\documentclass[aps,superscriptaddress,eqsecnum,nofootinbib,showpacs,preprintnumbers,twocolumn]{revtex4}

\pdfoutput=1
\usepackage{graphicx,epsfig}
\usepackage{amsmath}
\usepackage {amssymb}
\usepackage {longtable}
\usepackage{multirow}

\usepackage{dcolumn}
\usepackage{bm}
\usepackage{amsfonts}
\usepackage{subfigure}
\usepackage{color}

\usepackage{relsize}

\def\be{\begin{equation}}
\def\ee{\end{equation}}
\def\beq{\begin{eqnarray}}
\def\eeq{\end{eqnarray}}

\begin{document}

\title{Stability of Kerr black holes in generalized hybrid metric-Palatini gravity}

\author{Jo\~{a}o Lu\'{i}s Rosa}
\email{joaoluis92@gmail.com}
\affiliation{Centro de Astrof\'isica e Gravita\c c\~ao - CENTRA,
Departamento de F\'isica,
Instituto Superior T\'{e}cnico - IST,
Universidade de Lisboa - UL,
Avenida Rovisco Pais 1, 1049-001 Lisbon, Portugal}

\author{Jos\'{e} P. S. Lemos}
\email{joselemos@ist.utl.pt}
\affiliation{Centro de Astrof\'isica e Gravita\c c\~ao - CENTRA,
Departamento de F\'isica,
Instituto Superior T\'{e}cnico - IST,
Universidade de Lisboa - UL,
Avenida Rovisco Pais 1, 1049-001 Lisbon, Portugal}

\author{Francisco S. N. Lobo}
\email{fslobo@fc.ul.pt}
\affiliation{Instituto de Astrof\'{\i}sica e Ci\^{e}ncias do
Espa\c{c}o - IA, Departamento de F\'isica,
Faculdade de Ci\^encias - FC, Universidade de Lisboa - UL,
Campo Grande, 1749-016 Lisbon, Portugal}


\begin{abstract} 

It is shown that the Kerr solution exists in the generalized hybrid
metric-Palatini gravity theory and that for certain choices of the
function $f(R,\mathcal R)$ that characterizes the theory,
the Kerr solution can be stable against perturbations on the
scalar degree of freedom of the theory.
We start by verifying which are the most general conditions on the
function $f(R,\mathcal R)$ that allow for the
general relativistic Kerr solution to also be
a solution of this theory.
We perform a scalar perturbation
in the trace of the metric tensor, which in turn imposes a
perturbation in both the Ricci and Palatini scalar curvatures. To
first order in the perturbation, the equations of motion, namely the
field equations and the equation that relates the Ricci and the
Palatini curvature scalars, can be rewritten in terms of a fourth-order wave
equation for the perturbation $\delta R$ which can be factorized into
two second-order massive wave equations for the same variable.
The usual ansatz and separation methods are applied and stability
bounds on the effective mass of the Ricci scalar perturbation are obtained.
These stability regimes are studied case by case and specific forms of
the function $f(R,\mathcal R)$ that allow for a stable Kerr 
solution to exist within the perturbation regime studied are obtained.

\end{abstract}


\maketitle

\section{Introduction}\label{introduction}

General relativity, a relativistic theory of gravitation, see
e.g. \cite{mtw}, has passed a great number of tests, from the
weak-field tests within the Solar System to strong-field tests that include
black holes and gravitational waves.  In a cosmological setting, one
needs to add to the ingredients of general relativity some form of
dark matter to deal with the large-scale structure of the Universe,
and to postulate a cosmological constant, or a variant of it, to
explain the acceleration of the Universe.

In alternative to general relativity plus dark matter and cosmological
constant package, one can use a modified theory of gravitation, an
extension of general relativity, by modifying the gravitational sector
of the theory.  In this way one can also address the structure and
dynamics of the known self-gravitating systems and account for the
Universe's self-accelerated cosmic expansion. In $f(R)$ gravity
\cite{modgrav,modgrav2,modgrav3,Capozziello:2011et}, it has been
established that both metric and Palatini \cite{Olmo:2011uz} versions
of these theories have interesting features but also manifest severe
and different downsides. To overcome these problems, a hybrid
combination of theories, containing elements from both formalisms,
turns out to be fruitful in accounting for the observed phenomenology
and in addition is able to avoid some drawbacks of the original
approaches. This approach is known as the hybrid metric-Palatini
gravity \cite{harko,capozziello}. The action that describes this
theory is obtained from the usual Einstein-Hilbert action $R$ by the
addition of a function $f(\mathcal R)$, where $\mathcal R$ is a
curvature scalar defined in terms of an independent connection
$\hat\Gamma$. In this theory, the metric and the affine connection are
considered to be independent degrees of freedom, therefore combining
both the metric and the Palatini formalisms into a new modified
gravity. This theory was shown to be very successful in accounting for
observed phenomena in cosmological \cite{capozziello2} and galactic
dynamics \cite{capozziello3,Capozziello:2012qt}, leaving the Solar
System constraints unaffected \cite{harko}.  For a comprehensive and
extensive review on the hybrid metric-Palatini theory see
\cite{BookHarkoLobo}.

The generalized hybrid metric-Palatini (GHMP) gravity arises as a
natural outcome of the hybrid metric-Palatini gravity, where the
action $R+f(\mathcal R)$ is replaced by a general function
$f(R,\mathcal R)$ of both the Ricci and Palatini scalar curvatures
\cite{tamanini}. This theory was studied in the context of cosmology,
both with dynamical systems methods \cite{tamanini} and with
reconstruction techniques \cite{rosa1}, for which it was shown, among
other behaviors, that exponentially expanding cosmological models
exist even when the matter distribution is not purely vacuum. Also,
asymptotically anti-de Sitter
wormhole solutions with thin shells that satisfy
the null energy condition for the whole spacetime were obtained in
this theory \cite{rosa2}.

General relativity has produced as solutions, the static, i.e.,
Schwarzschild, black hole, and the rotating, i.e. Kerr
\cite{Kerr:1963ud}, black hole, which mirror the observed rotating
astrophysical black holes. As physically realistic objects, Kerr black
holes must be stable against exterior perturbations. Within general
relativity, the stability of Kerr black holes has been studied for
scalar, vectorial and tensorial perturbations. For massless
perturbations, the Kerr black hole was shown to be stable
\cite{teukolsky1974,chandrabook}.  For massive perturbations the issue
is more subtle, see e.g.~\cite{beyer}.  Moreover, for massive scalar,
vectorial and tensor perturbations, the confinement of superradiant
modes can lead to an amplification of the perturbation ad infinitum,
giving rise to instabilities such as the black hole bomb
\cite{press,lemos1}.

In $f(R)$ gravity black hole solutions and perturbations have also
been analyzed. An initial effort has been to reproduce and study
within $f(R)$ gravity the Schwarzschild and Kerr solutions of general
relativity. This theory was motivated to understand the acceleration
of the Universe in a natural way, and thus in principle, it contains
in it some form of a cosmological constant, meaning that the spacetime
is asymptotically de Sitter.  However, the Schwarzschild and Kerr
solutions are asymptotically flat rather than asymptotically de
Sitter, the rationale for using those, is that, as a first
approximation, locally, the influence of the cosmological constant
term is negligible and thus consideration of the Schwarzschild or Kerr
solutions is justified, besides being more simple.  Thus, confining to
Schwarzschild or Kerr, perturbation analyses of those solutions have
been performed within $f(R)$ gravity.  For instance, the stability of
the $f(R)$ Schwarzschild black hole in $f(R)$ theory was investigated
in its scalar-tensor representation by introducing two auxiliary
scalars \cite{myung1}. It was shown that the curvature scalar becomes
a scalaron, so that the linearized equations are second order and in
addition are the same equations as for the massive Brans-Dicke
theory. Furthermore, it was proved that the $f(R)$ black hole solution
is stable against external perturbations if the scalaron does not have
a tachyonic mass. The analysis was even extended to include the
stability of the Schwarzschild-AdS black hole in $f(R)$
theories with a negative cosmological constant \cite{Moon:2011sz} with
the conclusion that stable solutions against external perturbations
exist if the scalaron is again free from tachyons. The stability of
the Schwarzschild black hole was also analyzed in several extensions
of $f(R)$ gravity \cite{Moon:2011fw,Myung:2013doa,Myung:2017qtc}.  The
study of the stability of the Kerr solution in $f(R)$ gravity has been
studied in \cite{Myung:2011we,myung2}, where it has been proved that
it is unstable due to the fact that the perturbation equation for the
massive spin-0 graviton in this theory, or equivalently the perturbed
Ricci scalar, is analogous to a Klein-Gordon equation for a massive
scalar field in general relativity which has been intensively studied
and showed to be unstable.

In GHMP gravity it is also important to analyze black holes and their
stability. Again, the Schwarzschild and Kerr solutions are useful in
this theory.  The logic to study these solutions in $f(R,\mathcal R)$
gravity is the same as that used in $f(R)$, namely, although GHMP
gravity was motivated to understand the acceleration of the Universe
having some form of a cosmological constant, locally one can argue
that the influence of it is negligible and thus the use of the
Schwarzschild and Kerr solutions, rather than the asymptotically de
Sitter counterparts, is justified.  Since Kerr black holes are stable
within general relativity it is of interest to know whether those
black holes exist or not as solutions of the GHMP gravity and, in the
case that the answer is positive, it is important to perform
a stability analysis of the black holes themselves with the theory.
A first step in that direction is to understand
the perturbations in both the Ricci and Palatini scalar
curvatures of $f(R,\mathcal R)$ gravity within this setting
and
to work out for which
choices of the function $f(R,\mathcal R)$
Kerr black holes are stable
to those perturbations. This is what we set out to do here.

The paper is organized as follows. In Sec.~\ref{sec1}, we introduce
the action of the GHMP gravity and compute the respective equations of
motion.  In Sec.~\ref{sec2}, we start with a general form of the
function $f$ that guarantees that constant Ricci scalar $R$ solutions
exist in the GHMP gravity, and then choose the specific case of the
Kerr metric to compute perturbations to the massive spin-0 degree of
freedom.  In Sec.~\ref{sec3}, we compute the stability regimes of the
perturbations and the forms of the function $f$ that allow for these
regimes to be attained. In Sec.~\ref{conclusions}, we conclude.

\section{Action and field equations of the GHMP gravity}\label{sec1}

Consider the action $S$ of the GHMP gravity given by
\be\label{genac}
S=\frac{1}{2\kappa^2}\int_\Omega\sqrt{-g}f(R,{\cal R})d^4x+S_m,
\ee
where $\kappa\equiv 8\pi G$,
$G$ is the gravitational constant, $\Omega$ is the spacetime volume
and $d^4x$ its volume element, 
$g$ is the determinant of the spacetime
metric $g_{ab}$, 
$R$ is the metric Ricci scalar,
$\mathcal{R}\equiv\mathcal{R}^{ab}g_{ab}$ is the Palatini Ricci scalar,
where the Palatini Ricci tensor is defined in terms of an independent
connection $\hat\Gamma^c_{ab}$ as, 
\be
\mathcal{R}_{ab}=\partial_c
\hat\Gamma^c_{ab}-\partial_b\hat\Gamma^c_{ac}+\hat\Gamma^c_{cd}
\hat\Gamma^d_{ab}-\hat\Gamma^c_{ad}\hat\Gamma^d_{cb}\,,
\ee
$f(R,\cal{R})$ is a well-behaved
function of $R$ and $\cal{R}$, and
$S_m$ is the matter action defined as $S_m=\int d^4x\sqrt{-g}\;{\cal L}_m$,
where ${\cal L}_m$ is the matter Lagrangian density
considered minimally coupled to the metric $g_{ab}$. We
set the speed of light to one, $c=1$. 
Equation \eqref{genac} is the geometrical representation of the GHMP
gravity.
An equivalent scalar-tensor representation of the theory with
two scalar fields is possible to obtain with the help of auxiliary
scalar fields, see Appendix~\ref{scalarrep}.

Variation of the action  (\ref{genac})
with respect to the metric $g_{ab}$ yields the following equation of motion,
\beq\label{genfield}
\frac{\partial f}{\partial R}R_{ab}+
\frac{\partial f}{\partial \mathcal{R}}\mathcal{R}_{ab}-
\frac{1}{2}g_{ab}f(R,\cal{R})
   \nonumber \\
-\left(\nabla_a\nabla_b-g_{ab}\Box\right)
\frac{\partial f}{\partial R}=\kappa^2 T_{ab},
\eeq
where $\nabla_a$ is the covariant derivative and
$\Box=\nabla^a\nabla_a$ is the d'Alembertian
operator, both with respect to $g_{ab}$, and
$T_{ab}$ is the stress-energy tensor defined in the usual manner as
\begin{equation}
T_{ab}=-\frac{2}{\sqrt{-g}}\frac{\delta(\sqrt{-g}\,{\cal
L}_m)}{\delta(g^{ab})}\,.
 \label{defSET}
\end{equation}

Varying the action (\ref{genac}) with respect to the
independent connection $\hat\Gamma^c_{ab}$
provides the following relationship,
\be
\hat\nabla_c\left(\sqrt{-g}\,
\frac{\partial f}{\partial \cal{R}}g^{ab}\right)=0 \,,
\label{eqvar1}
\ee
where $\hat\nabla_a$ is the covariant derivative with respect to the
connection $\hat\Gamma^c_{ab}$.
Now recalling that $\sqrt{-g}$ is a scalar density of weight 1, we
have that $\hat\nabla_c \sqrt{-g}=0$ and so Eq.~(\ref{eqvar1})
simplifies to $\hat\nabla_c\left(\frac{\partial f}{\partial
\cal{R}}g^{ab}\right)=0$.  This means that there exists a new metric
$h_{ab}$ defined as
\be
h_{ab}=g_{ab} \frac{\partial f}{\partial \cal{R}}
\label{hab}
\ee
such that the connection $\hat\Gamma^a_{bc}$ is the
Levi-Civita connection for this metric, i.e.
\be
\hat\Gamma^a_{bc}=\frac{1}{2}h^{ad}\left(\partial_b h_{dc}+\partial_c
h_{bd}-\partial_d h_{bc}\right)\,,
\ee
where $\partial_a$ denotes a partial derivative.  Note also from
Eq.~(\ref{hab}) that $h_{ab}$ is conformally related to $g_{ab}$
through the conformal factor ${\partial f}/{\partial \cal{R}}$. This
result implies that the two Ricci tensors $R_{ab}$ and $\mathcal
R_{ab}$, that we assumed to be independent at first, are actually
related to each other by
\be\label{riccirel}
\mathcal R_{ab}=R_{ab}-\frac{1}{f_\mathcal R}\left(\nabla_a\nabla_b+
\frac{1}{2}g_{ab}\Box\right)f_\mathcal R+\frac{3}
{2f_\mathcal R^2}\partial_a
f_\mathcal R\partial_b f_\mathcal R,
\ee
where the subscripts $R$ and $\mathcal R$ denote derivatives of the
function $f$ with respect to either $R$ and $\mathcal R$,
respectively. Note that we shall be working with forms of the function
$f$ that satisfy the Schwartz theorem, which means that its crossed
derivatives are the same, i.e., $f_{R\mathcal R}=f_{\mathcal R R}$. We
therefore have a system of two independent equations of motion,
Eqs.~\eqref{genfield} and~\eqref{riccirel}, the latter being
equivalent to Eq.~\eqref{eqvar1}.

\section{Perturbations in GHMP of general relativity solutions
with $R_{ab}=0$}\label{sec2}

\subsection{General conditions on the function $f(R,\mathcal R)$}

In this section we assume a
general form for the
function $f(R,\mathcal R)$ that guarantees that 
general relativity solutions with $R_{ab}=0$, such as the Schwarzschild
and Kerr solutions, are also solutions of the GHMP theory.

To do so, let us assume two very general conditions for the function
$f(R,\mathcal R)$.
First, consider that the function $f$ is analytical in both $R$
and $\mathcal R$ around a point $\left\{0,\mathcal R_0\right\}$, where
$\mathcal R_0$ is a constant, and therefore can be expanded in a
Taylor series of the form
\beq\label{function}
f(R,\mathcal R)=\sum_{\left\{n,m\right\}=0}^\infty
\frac{\partial^{\left(n+m\right)}f
\left(0,\mathcal R_0\right)}{\partial^n R\ \partial^m\mathcal R} \times
	\nonumber \\
\times \frac{R^n}{n!}\frac{\left(\mathcal R-\mathcal R_0\right)^m}{m!}.
\eeq
Second, impose that the function $f$ has a zero at the point
where we perform the Taylor series expansion, that is
\beq\label{function2}
f\left(0,\mathcal R_0\right)=0.
\eeq

We now show that for a function $f$ that satisfies these
two conditions it is always possible for a general relativity
solution with $R_{ab}=0$ and so
$R=0$ to be also a solution in the GHMP gravity. To
start with, let $X$ denote $R$, $\cal R$, or any
combination of the form $R\cal R$, and so on, and
let $f_X$ denote the derivative of $f$ with respect
to $X$.
Then, the derivatives
of the functions $f_X$ with respect to the coordinates $x^a$ can be
written as functions of the derivatives of $R$ and $\mathcal R$ by
making use of the chain rule, from which we obtain
\be\label{c1}
\partial_a f_X=f_{XR}\partial_a R + f_{X\mathcal R}\partial_a\mathcal R,
\ee
which also allow us to write the terms $\nabla_a\nabla_b f_X$ and
$\Box f_X$ as functions of $R$ and $\mathcal R$ as
\beq\label{c2}
\nabla_a\nabla_bf_X=f_{XRR}\nabla_aR\nabla_bR+
f_{X\mathcal R\mathcal R}
\nabla_a\mathcal R\nabla_b\mathcal R +
	\nonumber \\
2f_{XR\mathcal R}\nabla_{(a}R\nabla_{b)}
\mathcal R+f_{XR}\nabla_a\nabla_bR+
f_{X\mathcal R}\nabla_a\nabla_b\mathcal R,
\eeq
where indices within parentheses are symmetrized, and
\beq\label{e222}
\Box
f_X=g^{ab}\nabla_a\nabla_bf_X\,.
\eeq
Now, let us first use
Eq.~\eqref{riccirel} to eliminate the term $\mathcal R_{ab}$ in
Eq.~\eqref{genfield}, from which we get
$
\left(f_R+f_\mathcal R\right)R_{ab}-\left(\nabla_a\nabla_b+
\frac{1}{2}g_{ab}\Box\right)f_\mathcal R+\frac{3}{2f_\mathcal R}
\partial_a f_\mathcal R\partial_b f_\mathcal R
   -\frac{1}{2}g_{ab}f-\left(\nabla_a\nabla_b-g_{ab}\Box\right)f_R
   =\kappa^2 T_{ab}
   $.
We want vacuum solutions of the 
GHMP theory and so we further assume $T_{ab}=0$.
Since we are also assuming from the start
that $R_{ab}=0$, this latter equation turns into
\beq\label{genfield2}
&&-\left(\nabla_a\nabla_b+\frac{1}{2}g_{ab}\Box\right)
f_\mathcal R+\frac{3}{2f_\mathcal R}\partial_a
f_\mathcal R\partial_b f_\mathcal R
   \nonumber \\
&&-\left(\nabla_a\nabla_b-g_{ab}\Box\right)f_R=0\,,
\eeq
where the expansions given in Eqs.~(\ref{c1}) and~(\ref{c2})
could have been inserted,  but we have not written the
final result due to its length.~Equation~(\ref{genfield2})
is a partial differential equation for $\mathcal
R$ that in principle cannot be solved until we choose a particular
form for the function $f$.  However, notice that if $\mathcal
R=\mathcal R_0$, where $\mathcal R_0$ is a constant, then
Eq.~(\ref{genfield2}) is identically zero upon using Eqs.~(\ref{c1})
and~(\ref{c2}), with the first assumption, i.e., Eq.~(\ref{function}),
guaranteeing that all the terms in Eq.~(\ref{genfield2}) are finite at
$R=0$ and ${\cal R}={\cal R}_0$.  We then take the particular solution
$\mathcal R=\mathcal R_0$.  Finally, tracing Eq.~\eqref{riccirel},
assuming $R=0$ and using the solution $\mathcal R=\mathcal R_0$ from
the previous equation, we obtain directly that $\mathcal R_0=0$.  Thus
solutions of general relativity with $R_{ab}=0$ 
are also solutions of
GHMP for which $R_{ab}=0$ and so $R=0$, and ${\cal R}=0$.
This result is consistent with the fact that we have chosen a specific
value for both $R$ and $\mathcal R$ in the previous paragraph, which
implies that the conformal factor between the metrics $g_{ab}$ and
$h_{ab}$, given by $f_\mathcal R$, is constant, the two metrics thus
have the same Ricci tensor, and so $\mathcal R=g^{ab}\mathcal
R_{ab}=g^{ab}R_{ab}=R$. Note that the field equation and the relation
between the scalar curvatures are both partial differential equations,
and therefore their solutions are not unique. We choose this
particular solution because it allows us to perform the following
analysis without specifying a form for the function $f$ besides the
two assumptions already made.

Thus,
we will work with the solutions 
\be\label{rab=0r=0}
R_{ab}=0\,,\quad R=0\,,
\ee
and 
\be\label{calr=0}
{\cal R}=0\,,
\ee
of the GHMP theory.

\subsection{Metric perturbations and linearized equations of motion}

Let us now consider a perturbation $\delta g_{ab}$ in the background
metric $\bar{g}_{ab}$, such that the new metric can be written as
\be\label{pert}
g_{ab}=\bar{g}_{ab}+\epsilon\delta g_{ab}, 
\ee
where $\epsilon$ is a small parameter. A bar here represents
unperturbed quantities. This perturbation in the metric induces a
perturbation in the Ricci tensor and Ricci scalar
of the form 
\be\label{pertriccitensor}
R_{ab}={\bar{R}_{ab}}+\epsilon\delta R_{ab}\,, 
\ee
\be\label{pertricciscalar}
R=\bar{R}+\epsilon\delta R\,,
\ee
respectively.
Through the definitions of 
$R_{ab}$ in terms of  $g_{ab}$ and its derivatives,
the perturbations $\delta
R_{ab}$ and $\delta R$ can be written in terms of $\delta g_{ab}$
and its derivatives as
\be\label{pertriccitensor2}
\delta R_{ab}=\frac{1}{2}\left(2\nabla^c\nabla_{(a}\delta
g_{b)c}-\Box\delta g_{ab}-\nabla_a\nabla_b\delta g\right)\,,
\ee
\be\label{pertricciscalar2}
\delta R=\nabla_a\nabla_b\delta g^{ab}-\Box\delta g\,,
\ee
where the parentheses in the indices denote index symmetrization and
$g$ is the trace of $g_{ab}$. Note that due to the conformal relation
between the metrics $g_{ab}$ and $h_{ab}$, a perturbation in the
former induces a perturbation in the latter, and thus both the
Palatini Ricci tensor $\mathcal R_{ab}$ and the Palatini scalar
$\mathcal R$ will also be written in terms of perturbations of the
form $\mathcal R_{ab}=\bar{\mathcal R}_{ab}+\epsilon\delta \mathcal
R_{ab}$ and $\mathcal R=\bar{\mathcal R}+\epsilon\delta \mathcal R$,
respectively.
The relation between the perturbations of the Palatini
tensor and scalar, $\delta \mathcal R_{ab}$ and $\delta\mathcal R$,
respectively,
and the perturbations of the Ricci tensor and scalar via
Eq.~\eqref{riccirel} perturbed to first order can be worked
out, as we shall see in a
moment.

Since the unperturbed quantities $\bar{R}$ and $\bar{\mathcal R}$
vanish in the solutions we are considering, see Eqs.~(\ref{rab=0r=0})
and~(\ref{calr=0}),
the
function $f$
and its derivatives $f_X$
can be expanded to 
first order in $\epsilon$ as
\be\label{expf}
f=\epsilon \left(\bar f_{R}\delta R+ \bar
f_{\mathcal R}\delta \mathcal R\right).
\ee
\be\label{expfx}
f_X=\bar f_X+\epsilon \left(\bar f_{XR}\delta R+ \bar
f_{X\mathcal R}\delta \mathcal R\right)\,.
\ee
respectively, where in Eq.~(\ref{expf})
we have used $\bar{f}(0,0)=0$, see Eq.~(\ref{function2})
with ${\mathcal R}_0=0$.
The expansions~(\ref{expf})
and~(\ref{expfx}) can also be
achieved using 
Eq.~(\ref{function}) with
$f(R,{\cal R})=
f( \bar{R}+\delta R,\bar{\cal{R}}+\delta\cal{R})$.
Note that the barred functions are constants,
because they represent
the coefficients of the Taylor expansion of the unperturbed function
$f$, and therefore they can be taken out of the derivative operators
unchanged, e.g. $\partial_a f=\epsilon\left(\bar f_R \partial_a \delta
R+\bar f_\mathcal R\partial_a\delta\mathcal R\right)$. To simplify the
notation, from now on we shall drop the bars, and any term containing
the function $f$ and its derivatives is to be considered as a constant.
In the 
scalar-tensor representation of the theory with
two scalar fields 
it can be shown that the perturbation analysis
remains the same, see Appendix~\ref{scalarrep} for more
details.

The equations of  motion~\eqref{genfield} and
\eqref{riccirel} then become, in vacuum and to first
order in $\epsilon$,
\beq
&& f_R\delta R_{ab}+f_{\mathcal R}\delta\mathcal R_{ab}-
\frac{1}{2}\bar g_{ab}\left(f_R\delta R+f_{\mathcal R}
\delta \mathcal R\right)-\nonumber\\
&& -\left(\nabla_a\nabla_b-g_{ab}\Box\right)
\left(f_{R R}\delta R+f_{R \mathcal R}\delta\mathcal R\right)=0\,,
\label{perturbfield}
\eeq
\beq
&&\delta\mathcal R_{ab}=\delta R_{ab}- \label{perturbricci}\\
&&-\frac{1}{f_{\mathcal R}}\left(\nabla_a\nabla_b+
\frac{1}{2}\bar g_{ab}\Box\right)\left(f_{\mathcal R R}
\delta R+f_{\mathcal R \mathcal R}\delta\mathcal R\right)\,,\nonumber
\eeq
respectively. These equations are fourth-order equations in the metric
perturbation $\delta g_{ab}$, and difficult to handle.

However, a system of equations for $\delta R$ and
$\delta\mathcal R$ can be obtained by taking the trace of
Eqs.~\eqref{perturbfield} and \eqref{perturbricci}, and
the perturbation analysis of this sector is simpler and can be
dealt with. This approach is
well motivated: similarly to the $f(R)$ theories of
gravity, the GHMP theory presents three degrees of freedom without
ghosts, two for massless spin-2 gravitons, and one for a massive
spin-0 scalar graviton. One might think that there are two scalar
degrees of freedom corresponding to both $f_R$ and $f_{\mathcal R}$,
but these actually correspond to the same degree of freedom due to
their conformal relation expressed by the trace of
Eq.~\eqref{riccirel}. Now, the scalar degree of freedom is well
described by the trace $\delta g$.
Using the Lorenz gauge, i.e.,
$\nabla_b\delta g^{ab}=(1/2)\nabla^a\delta g$,
Eq.~\eqref{pertricciscalar2} turns into
$\delta R=\nabla_a\nabla_b\delta g^{ab}-\Box\delta g=
-\frac{1}{2}\Box\delta g$.
So, under this gauge,
the perturbation $\delta R$ is directly related
to $\delta g$ which
represents the massive spin-0 degree of freedom of the theory.
We 
restrict ourselves to the study of the massive scalar degree of
freedom of the GHMP theory by the analysis of the perturbation $\delta
R$, i.e., we will study stability
against scalar mode perturbations. 

To obtain an equation for the perturbation in the Ricci scalar $\delta
R$ we shall work with the traces of Eqs.~\eqref{perturbfield}
and~\eqref{perturbricci}. These equations become
\be\label{eqsys1}
f_R\delta R+f_\mathcal R\delta\mathcal R-3f_{RR}\Box\delta R-
3f_{R\mathcal R}\Box\delta\mathcal R=0,
\ee
\be\label{eqsys2}
\delta\mathcal R=\delta R-\frac{3}{f_\mathcal R}
\left(f_{\mathcal R\mathcal R}\Box\delta\mathcal R+
f_{\mathcal R R}\Box\delta R\right),
\ee
respectively, where we used $T_{ab}=0$ and $\bar
f=f\left(0,0\right)=0$. Note that the perturbations $\delta R$ and
$\delta\mathcal R$ cannot be equal. If they were, then one of the
equations above would immediately set $f(R,\mathcal
R)=f(R-\mathcal R)$, and thus the perturbations
would cancel completely in the other equation and we would
obtain an
identity. This is not a feature of the first-order expansion, for it
can be shown with some care 
that for any order in $\epsilon$ that we choose, if
$f(R,\mathcal R)=f(R-\mathcal R)$ then the
perturbations cancel identically in these two equations.

\begin{figure}[h]
\includegraphics[scale=0.5]{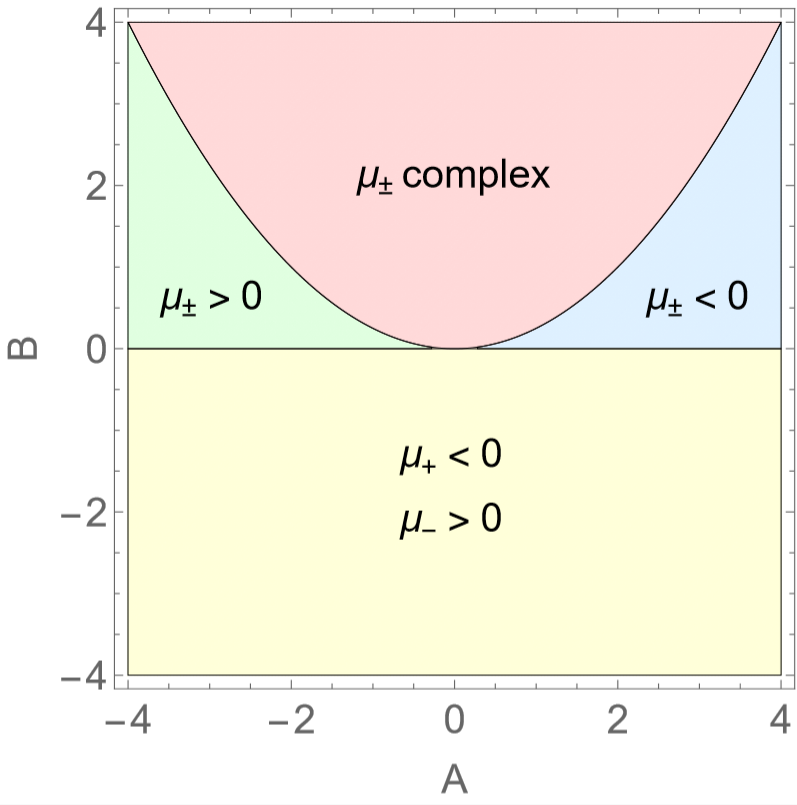}
\caption{The properties of the masses $\mu_\pm$
in the parameter space $(A,B)$ are displayed.}
\label{fig:masses}
\end{figure}

Equations~\eqref{eqsys1} and~\eqref{eqsys2} can both be rewritten in the
form $\left(\Box+a_1\right)\delta
R=a_2\left(\Box+a_3\right)\delta\mathcal R$, where
$a_1$, $a_2$, and $a_3$
are constants
that depend only on the values of $f_X$ and that are different
for both equations. To obtain an equation that depends only on $\delta
R$, we proceed as follows. First, we solve Eq.~\eqref{eqsys2} with
respect to $\Box\delta\mathcal R$ and we replace it in
Eq.~\eqref{eqsys1} to obtain an equation of the form
$\left(\Box+b_1\right)\delta R=b_2\delta\mathcal R$, where $b_1$
and $b_2$ are constants. Second,
we solve Eq.~\eqref{eqsys2} with respect
to $\Box\delta R$ and insert the result into Eq.~\eqref{eqsys1} to
obtain an equation of the form $\left(\Box+c_1\right)\delta \mathcal
R=c_2\delta R$, where $c_1$ and $c_2$ are constants.
Third,  we
use the first of these two equations to replace the term depending
on $\delta\mathcal R$ in the second equation. The resultant equation is
\be\label{box2}
\Box^2\delta R+A\Box\delta R+B\delta R=0,
\ee
where the constants $A$ and $B$ are given in terms of the
background quantities
$f_X$ as
\beq
A&=&\frac{f_Rf_{\mathcal R\mathcal R}-2f_\mathcal
Rf_{\mathcal R R}-f_\mathcal Rf_{RR}}{3
\left(f_{\mathcal R R}^2-f_{RR}f_{\mathcal
R\mathcal R}\right)},\label{ABa} \\
B&=&\frac{f_\mathcal R\left(f_\mathcal R+
f_R\right)}{9\left(f_{\mathcal R R}^2-
f_{RR}f_{\mathcal R\mathcal R}\right)}.\label{AB}
\eeq
Note that Eq.~\eqref{box2} is a fourth-order equation in the perturbation
$\delta R$. However, since $A$ and $B$ are constants, it is possible to
factorize Eq.~\eqref{box2} into
\be\label{2box}
\left(\Box-\mu_+^2\right)\left(\Box-\mu_-^2\right)\delta R=0,
\ee
where the constants $\mu_\pm^2$ can be expressed in terms of the
constants $A$ and $B$ as
\beq\label{mus}
\mu_+^2&=-\frac{1}{2}\left(A+\sqrt{A^2-4B}\right)\,,\nonumber\\
\mu_-^2&=-\frac{1}{2}\left(A-\sqrt{A^2-4B}\right)\,,
\eeq
with their main properties  in terms
of the parameters $A$ and $B$ being plotted in
Fig.~\ref{fig:masses}. Note that, since
$\mu_\pm^2$ are
constants, the terms
$\left(\Box-\mu_+^2\right)$ and
$\left(\Box-\mu_-^2\right)$
commute in Eq.~\eqref{2box}, and so
we can reduce Eq.~\eqref{2box}
into a set of two equations of the form
\be\label{3box}
\left(\Box-\mu_+^2\right)\delta R=0\,,\quad
\left(\Box-\mu_-^2\right)\delta R=0\,,
\ee
which are of the form of a
Klein-Gordon equation for a scalar field where the constants
$\mu_\pm^2$ take the role of the field's mass.
Thus, the scalar mode of the perturbation
is a massive mode. 

We can state in brief,
that when we perturb the metric tensor, the equation that describes the
perturbation in the Ricci scalar is a fourth-order massive wave equation
with two different masses. However, since the Ricci
scalar perturbation depends on second-order
derivatives of the metric perturbation, we would
expect to be confronted
with a six-order differential equation involving $\delta g_{ab}$, with a
very complicated and untreatable form. The use of the Lorenz
gauge is what enables one to reduce
this equation to a fourth-order equation for a
massive spin-0 degree of freedom. This fourth-order equation can be
factorized into two commutative second-order equations of the form of a
massive
Klein-Gordon in general
relativity. One can now apply the usual separation methods to expand
the perturbation into spheroidal harmonics and a radial wavefunction.
If wished one can use
numerical integration techniques to compute the quasibound
state frequencies.

\section{The Kerr solution in GHMP: Equations, superradiant
instabilities and stability regimes} \label{sec3}

\subsection{Separability of the equations of motion}

\subsubsection{Separability of the equations of motion, quasibound
state}

Equations of the form (\ref{3box}) have been  studied and
are known to be separable for the Schwarzschild and Kerr metrics.  We
will be working with the Kerr metric, knowing that
the Schwarzschild
metric can be directly obtained from the Kerr
metric by taking the limit
where the angular momentum is equal to zero. The Kerr
metric in Boyer-Lindquist coordinates $(t,r,\theta,\phi)$
is given by
\beq
&&ds^2=-\left(1-\frac{2Mr}{\rho^2}\right)dt^2+\frac{\rho^2}{\Delta}
dr^2+\rho^2d\theta^2
	\nonumber \\
&&-\frac{4Mra\sin^2\theta}{\rho^2}dtd\phi
	\nonumber \\
&&	+\left(r^2+a^2+
\frac{2Mra^2\sin^2\theta}{\rho^2}\right)d\phi^2,
\eeq
with
\be
\Delta=r^2+a^2-2Mr, \quad \rho^2=r^2+
a^2\cos^2\theta,\quad a=\frac{J}{M},
\ee
where $M$ is the black hole mass and $J$ is the black hole angular
momentum. The event horizon of the Kerr black hole is at
the radius $r_+$ given by
\begin{equation}
r_+=M+\sqrt{M^2-a^2}\,.
\label{horizonradius}
\end{equation}

To study the separability of the equations of motion, we first note
that Eq.~\eqref{2box} is a fourth-order partial differential
equation (PDE) for $\delta R$, and should
therefore have four linearly independent solutions.
The equation $\left(\Box-\mu_+^2\right)\delta R=0$
has two solutions, one solution
corresponding to ingoing waves, the other corresponding to
outgoing waves. Also, the 
equation $\left(\Box-\mu_-^2\right)\delta R=0$
has two solutions, one solution
corresponding to ingoing waves, and the other corresponding to
outgoing waves.
To find these solutions
we
chose an ansatz of the form
\be\label{ansatz}
\delta R=\psi\left(r\right)S\left(\theta\right)\exp
\left(-i\omega t+im\phi\right), 
\ee
where $\psi\left(r\right)$ is the radial wavefunction, $\omega$ is the
wave angular frequency, $m$ is the azimuthal number, and
$S\left(\theta\right)$ are the scalar spheroidal harmonics.

Using this ansatz, we can separate each of the factors
$\left(\Box-\mu_\pm^2\right)\delta R=0$ into a radial and an angular
equation. The angular equation is given by
\beq
\left[\lambda-m^2+a^2\left(\omega^2-\mu_\pm^2\right)
\cos^2\theta\right]\sin^2\theta S\left(\theta\right)
	\nonumber\\
+\sin\theta\partial_\theta\left[\sin\theta
\partial_\theta S\left(\theta\right)\right]=0\,,
\label{spheroidal}
\eeq
where $\lambda=l\left(l+1\right)+ f\left(c\right)$,
$l$ is the angular momentum number,
$c=a^2\left(\omega^2-\mu_\pm^2\right)$ is
a constant,
and $f(c)$ is some function of $c$ that in the regime
we are working is negligible $f(c)={\mathcal O}(c)$ with
$c\ll 1$ as we will show.
In this case, thus, 
the spheroidal
harmonics can be approximated by the spherical harmonics, with
a constant of separation $\lambda=l\left(l+1\right)$.

Using Eq.~\eqref{spheroidal} in Eq.~\eqref{3box}
one finds a radial equation for the radial wave function
$\psi\left(r\right)$. To
find a more suitable way to write this radial equation, it is useful to
redefine
the radial coordinate $r$ and 
the radial wave function $\psi\left(r\right)$. Let us
define the tortoise coordinate $r_*$ and a new radial
wavefunction $u\left(r\right)$ as
\be\label{wave}
\frac{dr}{dr_*}=\frac{\Delta}{r^2+a^2},\qquad u
\left(r\right)=\sqrt{r^2+a^2}\,\psi\left(r\right),
\ee
so that the new radial equation can be written in the form of a wave
equation in the presence of a potential barrier as
\be\label{wave2}
\frac{d^2u}{dr_*^2}+\left[\omega^2-V\left(r\right)\right]u=0,
\ee
where the potential is given by
\beq\label{pot1}
V\left(r\right)&&=\frac{\Delta}{r^2+a^2}
\left[\frac{\Delta+\Delta'r}{\left(r^2+a^2\right)^2}
-\frac{3r^2\Delta}{\left(r^2+a^2\right)^3}+\frac{1}{r^2+a^2}\times
\right.
	\nonumber \\
&&\left.\left(\mu_\pm^2r^2-\omega^2a^2+\frac{4Mram\omega}{\Delta}
-\frac{m^2a^2}{\Delta}+\lambda\right)\right]\hskip-0.1cm.
\eeq
Equation \eqref{wave2} admits two solutions, one
corresponding to an ingoing wave and one to an
outgoing wave. Due to the complicated form of the potential
$V\left(r\right)$
of Eq.~\eqref{pot1}, Eq.~\eqref{wave2}
has no direct analytical solution and we
resort to solving the equation numerically.
For that we impose appropriate boundary conditions at the
horizon, where $r=r_+$ and
$r_*=-\infty$, and
at infinity, where $r=+\infty$
and $r_*=+\infty$.

\subsubsection{Quasibound
states}

At the horizon $r_+$ the potential in Eq.~\eqref{pot1}
takes the form $V\left(r_+\right)=\omega^2-
\left(\omega-m\Omega\right)^2$,
where $\Omega=\frac{a}{2Mr_+}$ is the angular velocity of the
horizon itself.
Thus, Eq.~\eqref{wave2} is
$\frac{d^2u}{dr_*^2}+\left(\omega-m\Omega\right)^2 u=0$.  The solution
is $u\left(r\right)=A_+
e^{i\left(\omega-m\Omega\right)r_*}+
B_+e^{-i\left(\omega-m\Omega\right)r_*}$,
for some constants of integration $A_+$ and $B_+$.
Since the horizon functions as a one-directional membrane, we want our
boundary condition at the horizon to be given by a purely ingoing
wave, i.e.,  at the horizon
there are no outgoing waves, so the corresponding $A_+$ is zero,
$A_+=0$. The solution is then $u\left(r\to
r_+\right)=B_+e^{-i\left(\omega-m\Omega\right)r_*}$.
At infinity the potential in Eq.~\eqref{pot1}
takes the form $V\left(\infty\right)=\mu_\pm^2$.
Thus, Eq.~\eqref{wave2} is
$\frac{d^2u}{dr_*^2}+\left(\omega^2-\mu_\pm^2\right)u=0$.
The solution
is $u\left(r\right)=A_\infty
e^{i\sqrt{\omega^2-\mu_\pm^2}r}
+B_\infty
e^{-i\sqrt{\omega^2-\mu_\pm^2}r}$,
for some constants of integration $A_\infty$ and $B_\infty$.
At infinity  we want the solution to decay
exponentially to give rise to a quasibound state, i.e., 
at infinity we want no waves and a decaying solution,
so $\omega^2<\mu_\pm^2$
and $B_\infty=0$.
The solution is then
$u\left(r\to\infty\right)=A_\infty e^{-\sqrt{\mu_\pm^2-\omega^2}r}$.
In brief, at the horizon and at
infinity the solutions are
\beq
u\left(r\to r_+\right)=B_+e^{-i\left(\omega-m\Omega\right)r_*},
	\nonumber \\
u\left(r\to\infty\right)=A_\infty e^{-\sqrt{\mu_\pm^2-\omega^2}r},
\label{boundary}
\eeq
respectively.

Finding the quasibound states consists of integrating the radial
Eq.~\eqref{wave} subjected to the boundary conditions in
Eq.~\eqref{boundary} and computing the roots for $\omega$. These roots
will be of the form $\omega=\omega_R+i\omega_I$,
with $\omega_R$ being the real part of the frequency
and $\omega_I$ its imaginary part. As can be seen from
Eq.~\eqref{ansatz}, if $\omega_I<0$ the perturbation decays
exponentially with time, but if $\omega_I>0$ the wavefunction grows
exponentially and at some later time
can no longer be considered a perturbation.  These
frequencies
have been calculated in several places \cite{teukolsky1974}
and we will not do it here. We want to
study the instability and the stability of Kerr black holes
in GHMP theory, so we proceed to
such an analysis.

\subsection{Superradiant stability regimes}
\subsubsection{General considerations about stability}

As explained, each of the terms $\left(\Box-\mu_\pm^2\right)\delta
R=0$, see Eq.~\eqref{3box},
gives rise to a set of two different solutions,
corresponding to an ingoing and an
outgoing wave. Since these terms commute in the full equation given
by Eq.~\eqref{2box}, the complete solution for this equation is
given by a linear combination of the two sets of solutions for each of
the $\mu_\pm^2$'s. Since Eq.~\eqref{2box} is a fourth-order equation,
these four solutions represent all the possible solutions for the
equation. As the masses $\mu_\pm$ are different in general,
the two sets
of solutions will form quasibound states for different ranges of the
angular frequency $\omega$.
Note that if one of the two sets of
solutions is unstable, then the entire solution will also be unstable,
even if the other set is stable.
The case $\mu_\pm=0$ is special in the sense that
the solutions will be decaying oscillating solutions
and so there are no quasibound states.
If the superradiant condition
$\omega<m\Omega$ is not satisfied the solution
will be automatically stable. Let us now
show that it is still possible to have
stability even if there is superradiance.
In this case
there are two ways the solution can be stable.

The first way to have stability even if there is superradiance
is to consider massless
perturbations, $\mu_\pm^2=0$.  In this case, there might be
superradiant modes, but quasibound states never form, and so clearly
the perturbation is stable.

The second way  to have stability even if there is superradiance
is to have a stable quasibound state. 
So, in this case the solution obeys the superradiant condition,
namely,
$\omega<m\Omega$. 
The solution has to have quasibound states, so the conditions 
$\mu_\pm>0$ and $\omega^2<\mu_\pm^2$ hold. Thus, 
we have 
\begin{equation}
\omega<{\rm min}(m\Omega,\mu_\pm)\,.
\label{supercon}
\end{equation}
Moreover, to have stable bound states
it is a sufficient condition
that $\mu_\pm$ obeys \cite{beyer}
\begin{equation}
\mu_\pm>\mu_c\,,\quad \mu_c=m\Omega\sqrt{1+\frac{2M}{r_+}}\,.
\label{muc}
\end{equation}

We can also achieve stability for a
combination of the two cases above, i.e., one of the masses might
vanish and the other might be in the range $\mu>\mu_c$.

Note that since $m$ is
an azimuthal number, it does not have an upper bound, and so one
could argue that for any constant value of $\mu_\pm^2$, there is
always a value of $m$ such that $\mu_\pm<\mu_c$. However, it
has been shown that superradiant instabilities are exponentially
suppressed for larger values of $m$. This implies that we can consider
an upper bound on $m$ for which the instability timescale is greater
than the age of the Universe, say $m^{\rm max}$, and only after
we choose an
appropriate value of $\mu_\pm$ that satisfies the inequality
$\mu_\pm>\mu_c^{\text{max}}$. This guarantees that even if the
instabilities occur, their effects would not be seen.

\subsubsection{Stability regimes: Sufficient conditions
on $f(R,\mathcal R)$}

\noindent{\it A. The case $\mu_\pm^2=0$:}

\noindent 
Let us start by studying the  case where the masses
$\mu_\pm^2$ vanish, $\mu_\pm^2=0$,
which implies that quasibound states can never
form and hence no instabilities can occur. From Eqs. \eqref{box2} and
\eqref{mus}, we verify that if both $A$ and $B$ vanish, then
Eq.~\eqref{2box} becomes simply $\Box^2\delta R=0$. This
corresponds to the
origin of the plot in Fig.~\ref{fig:masses}. If we can find a
form of the function $f(R,\mathcal R)$ such that both $A$
and $B$ vanish, then the Kerr solution will always be stable in this
$f(R,\mathcal R)$
theory.

To guarantee that none of the equations of motion diverge, we
need to guarantee that all the first and second derivatives of $f$,
i.e. $f_R$, $f_\mathcal R$, $f_{RR}$, $f_{\mathcal R\mathcal R}$,
$f_{R\mathcal R}$, are finite. On the other hand, the factors $A$
and $B$, given by Eqs.~\eqref{ABa} and~\eqref{AB}, will vanish if the
following conditions are satisfied,
$f_{R\mathcal R}^2-f_{RR}f_{\mathcal R\mathcal R}\neq 0$,
$f_Rf_{\mathcal R\mathcal R}-2f_\mathcal Rf_{R\mathcal R}-f_\mathcal R
f_{RR}=0$, and
$f_R+f_\mathcal R=0$.
Note that these conditions must be satisfied at $R=\mathcal
R=0$. There are many different functions $f$ that satisfy these
conditions. The simplest class of functions $f$ that satisfies these
conditions is
\be\label{fmu0}
f(R,\mathcal R)=\left(a_1+a_2R+a_3
\mathcal R\right)\left(R-\mathcal R\right)
\ee
where $a_1$, $a_2$, and $a_3$
are constants that must satisfy the constraint
$a_2\neq -a_3$. Any higher-order form of the function
$f(R,\mathcal R)$ obtained from Eq.~\eqref{fmu0} by adding
terms such as $R^3$ or $R^2\mathcal R$ will also have stable solutions
because all these extra terms vanish when we set $R=0$ and $\mathcal
R=0$ in Eqs. \eqref{ABa} and \eqref{AB}.

\vskip 0.3cm
\noindent {\it B. The case $\mu_-=0$ with $\mu_+>\mu_c$: }

\noindent 
Here we want $\mu_-=0$ with $\mu_+>\mu_c$.  As can be seen from
Eq.~\eqref{mus}, the only way for $\mu_-=0$ is to have $B=0,\, A>0$,
but these constraints impose that $\mu_+<0$,
see Fig.~\ref{fig:masses}, so $\mu_+$ can never be
greater than $\mu_c$.
So there are no forms of
$f(R,\mathcal R)$
for which the conditions 
$\mu_-=0$ with $\mu_+>\mu_c$ are satisfied.

\vskip 0.3cm
\noindent {\it C. The case $\mu_+=0$ with $\mu_->\mu_c$:}

\noindent 
Let us now set
$\mu_+=0$ by choosing $B=0, A<0$, and in this region we have
$\mu_->0$, see Fig.~\ref{fig:masses},
and we have to see whether
we can choose the function $f$ in such a way that
$\mu_->\mu_c$ or not.

As before, in order to avoid divergences in
the equations of motion we have to guarantee that all the first and
second derivatives of $f$, i.e. $f_R$, $f_\mathcal R$, $f_{RR}$,
$f_{\mathcal R\mathcal R}$, $f_{R\mathcal R}$, are finite, and the
extra constraints on the function $f$ such that $B=0$ and $A\neq 0$
are
$f_{R\mathcal R}^2-f_{RR}f_{\mathcal R\mathcal R}\neq 0$,
$f_Rf_{\mathcal R\mathcal R}-2f_\mathcal Rf_{R\mathcal R}$, and
$f_R+f_\mathcal R=0$.
These conditions must be satisfied at $R=\mathcal R=0$.
A simple class of functions $f$ that satisfies these
constraints is
\be\label{fmu1}
f(R,\mathcal R)=
a_1( R-\mathcal R)+
a_2R^2+
a_3\mathcal R^2+
a_4R
\mathcal R,
\ee
where  $a_1$, $a_2$, $a_3$, and $a_4$
are constants. This form of the function $f$
implies, by Eqs. \eqref{ABa}, \eqref{AB} and \eqref{mus}, that
$\mu_-$ and $A$
can be written as
\be\label{A1}
\mu_-=-A=\frac{2a_1\left(a_2+a_3+a_4\right)}{12 a_2a_3-3a_4^2}\,.
\ee
In order that  the solutions are stable we have to guarantee
that the $\mu_-$ of Eq.~(\ref{A1}) is greater than
$\mu_c$, i.e.,
$\frac{2a_1\left(a_2+a_3+a_4\right)}{12 a_2a_3-3a_4^2}>\mu_c$.
To obtain a finite $A$, and thus a finite $\mu_-$,
both the numerator and the denominator of
Eq.~\eqref{A1} must be $\neq 0$.
Now, let us try to find a specific combination of the constants
$a_1$, $a_2$, $a_3$, and $a_4$
such that $\mu_->\mu_c$
is satisfied. There are many combinations
that work, but let us take for example the case where $a_3$
and $a_4$ are set and verify if there is a value of $a_2$ that solves
the problem. Note that the denominator of Eq.~\eqref{A1} diverges to
$+\infty$ when we take the limit $a_2\to a_4^2/\left(4
a_3\right)$ from above. Also, if we choose both $a_3>0$ and $a_4>0$, then
$a_2>0$ and the numerator of Eq.~\eqref{A1} is positive in this
limit. This implies that we can always choose a finite value of
$a_2\gtrsim a_4^2/\left(4 a_3\right)$ arbitrarily close to
$a_4^2/\left(4 a_3\right)$ such that for any $m$ and $\Omega$ the
condition $\mu_->\mu_c$ is always satisfied.
Note that $m$
does not have an upper limit, but since  superradiant
instabilities are exponentially suppressed for larger values of $l$
and $m$, one has that
for $m \gg 1$ the effects of these instabilities are
negligible.
Again, any higher-order form of the function $f(R,\mathcal
R)$ will also have stable solutions because all these extra
terms vanish when we set $R=0$ and $\mathcal R=0$ in Eqs. \eqref{ABa}
and \eqref{AB}.

\vskip 0.3cm
\noindent {\it D. The case $\mu_\pm>\mu_c$:}

\noindent 
Finally, we turn to the case where both masses are
$\mu_\pm>\mu_c$. In this case we need
both $A$ and $B$ to be finite. Let us  analyze the regions of the
parameter space of $A$ and $B$ that allow for these solutions to
exist. From Eq.~\eqref{mus} and Fig.~\ref{fig:masses},
we can see that there are three regions
of the phase space that must be excluded: 1. Region $4B>A^2$, because
both $\mu_\pm^2$ are complex; 2. Region $B<0$, because $\mu_+$ is
always negative; 3. Region $B>0$ and $A>0$, because both $\mu_\pm^2$
are negative. We are then constrained to work in the region defined by
the three conditions $B>0$, $A<0$ and $A^2>4B$.

The approach to this problem is different from the previous ones. Let
us first study the structure of the $\mu_\pm^2$ as functions of $A$
and $B$ in Eq.~\eqref{mus}. In the limit of large $\mu_\pm^2$, we must
have $A\to -\infty$ like in the previous case. However, in this limit,
the quantity inside the square root is going to depend on how $B$ is
proportional to $A$. If $B\propto |A|^n$ with $n < 2$, then in this
limit $\sqrt{A^2-4B}\to |A|$ and $\mu_+ \to 0$, and we recover the
previous case. On the other hand, if $B\propto |A|^n$ with $n>2$, then
in this limit we eventually break the relation $A^2>4B$ and the
$\mu_\pm^2$ become complex. Therefore, we need a behavior of $B$ of
the form $B=CA^2$ for some constant $C$. Then, the constraints $B>0$
and $A^2>4B$ imply that the constant $C$ must be somewhere in the
region $0<C<1/4$. Inserting this form of $B$ into Eq.~\eqref{mus}
leads to the relation
$\mu_\pm^2=-(1/2)A\left(1\mp\sqrt{1-4C}\right)>0$, where the
inequality arises since we know that $A<0$ and $C<1/4$. At this point,
if we can choose a specific form of the function $f$ such that we can
make $A$ arbitrarily large, our problem is solved. Also note that in
the limit $C=0$ we recover the previous case {\it C} where $\mu_+=0$,
and in the limit $C=1/4$ we obtain $\mu_+=\mu_-$ which can be shown to
give $\mu_+=\mu_-=0$ recovering case {\it A}, i.e., $\mu_\pm^2=0$.

Consider now the most general form of the function $f$ that avoids
divergences in the equations of motion, i.e., a function $f$ for which
$f_R$, $f_\mathcal R$, $f_{RR}$, $f_{\mathcal R\mathcal R}$ and
$f_{R\mathcal R}$ are finite,
\be
f(R,\mathcal R)=a_1 R+a_2\mathcal R+
a_3R^2+a_4\mathcal R^2+a_5R\mathcal R,
\ee
where $a_1$, 
$a_2$, 
$a_3$,
$a_4$, and 
$a_5$, 
are constants assumed different from zero.
For this particular choice of $f$,
we see by Eq.~\eqref{AB} that $A$ diverges to $-\infty$ if the
numerator is positive and we take the limit $a_3\to a_5^2/\left(4
a_4\right)$ from above, or if the numerator is negative and we
take the limit
$a_3\to a_5^2/\left(4 a_4\right)$ from below.
However, this  is not
enough to conclude that we can make $\mu_\pm^2$ arbitrarily large. We
also need to verify that $B=CA^2$, with $0<C<1/4$, i.e., from
Eq.~\eqref{AB} we must have
$f_\mathcal R\left(f_\mathcal R+f_R\right)
\left(f_{\mathcal R R}^2-f_{RR}f_{\mathcal R\mathcal R}\right)=
	C\big(f_Rf_{\mathcal R\mathcal R}
-2f_\mathcal Rf_{\mathcal R R}-f_\mathcal Rf_{RR}\big)^2$
with 
$f_{\mathcal R R}^2-f_{RR}f_{\mathcal R\mathcal R}\neq 0$.
Finding the most general combinations of  $a_1$, 
$a_2$, 
$a_3$,
$a_4$, and 
$a_5$, for which the function
$f$ satisfies these constraints is a fine-tuning problem. To solve this
problem we proceed as follows. First we write $a_3=a_5^2/\left(4
a_4\right)+\epsilon$, for some $\epsilon\gtrsim 0$ that must be finite
but we can make it arbitrarily small. We also need to have $a_2\neq
a_1$ to guarantee that $B\neq 0$, so
it is better to redefine $a_2$ as an $a_6$ given by
$a_2\equiv-\left(a_6+1\right)a_1$, where $a_6\neq 0$. Inserting these
considerations into
$f_\mathcal R\left(f_\mathcal R+f_R\right)
\left(f_{\mathcal R R}^2-f_{RR}f_{\mathcal R\mathcal R}\right)=
	C\big(f_Rf_{\mathcal R\mathcal R}
-2f_\mathcal Rf_{\mathcal R R}-f_\mathcal Rf_{RR}\big)^2$,
we verify that $C$ is positive
in the small-$\epsilon$ limit only if the condition
$a_4a_6\left(1+a_6\right)<0$. This happens in the regimes $-1<a_6<0$
with $a_4>0$, $a_6<-1$ with $a_4<0$, and $a_6>0$ with $a_4<0$. As an
example, let us consider $a_4=1$, and then
$a_6=-1/2$ which corresponds to
the maximum of the polynomial $-a_6\left(a_6+1\right)$. Finally, we
have to guarantee that $A<0$ in this regime. Inserting these results
into Eq.~\eqref{ABa} we verify that $A<0$ requires that the quantities
$\left(a_5+1\right)$ and $a_1$ have the same sign. For simplicity, let
us take $a_1=a_5=1$.
So in this example we have $a_1=1$, 
$a_1=-\frac12$,
$a_3=\frac14+\epsilon$,
$a_4=1$, and 
$a_5=1$.
Note that other choices for the values of the
parameters  $a_1$, 
$a_2$, 
$a_3$,
$a_4$, and 
$a_5$, could also be made following the same reasoning.
Here, our aim is simply to provide an example of a combination that
works. We are thus left with
\beq
\mu_\pm^2=-\frac{A}{2}\left(1\mp\sqrt{1-4C}\right)\,,\quad\nonumber\\
A=-\frac{13+4\epsilon}{48\epsilon}\,,\quad
C=\frac{16\epsilon}{\left(13+4\epsilon\right)^2}\,,
\eeq
From these results, we verify that for any $\epsilon>0$ we have,
$0<C<1/4$, and also that in the limit $\epsilon\to 0$ we have $A\to
-\infty$ and thus $\mu^2_\pm\to +\infty$.
We can thus
consider $\epsilon$ arbitrarily small and force
$\mu^2_\pm>\mu_c$ for any $m$ and $\Omega$. We note
that $m$ does not have an upper bound, but superradiant instabilities
are exponentially suppressed for large values of $m$ and we can neglect
their effects.
Again, any higher-order form of the
function $f(R,\mathcal R)$ will also work because the extra
terms vanish for $R=\mathcal R=0$.

\section{Conclusions}\label{conclusions}

Within GHMP with its generic function $f(R,\mathcal R)$, we
have shown that it is always possible to choose a specific value for
$\mathcal R$, namely $\mathcal R=\mathcal R_0$ for some solution in
general relativity with constant $R=0$ such that this solution is
also a solution for the GHMP gravity for any form of the function $f$
that satisfies two very general conditions: $f$ must be analytical in
the point $\{0,\mathcal R_0\}$, and $f$ must have a zero in the same
point, i.e., $f\left(0,\mathcal R_0\right)=0$.
Inserting this result into the field equations leads
to the conclusion that $\mathcal R_0=0$.
This result is in
agreement with the fact that for constant $R$ and $\mathcal R$, the
conformal factor between the metrics $g_{ab}$ and $h_{ab}$, which is
given by $f_\mathcal R$, is constant and therefore both metrics
$g_{ab}$ and $h_{ab}$ must have the same Ricci tensor.

We have extended the scrutiny of the GHMP gravity by studying which
functions $f(R,\mathcal R)$ yield stability against scalar
perturbations of Kerr black hole solutions.  The stability of the Kerr
metric against superradiant instabilities is dictated by two
conditions: either the masses of the perturbations vanish,
$\mu_\pm=0$,
or the masses of these perturbations exceed a critical value
$\mu_\pm>\mu_c$.  We have shown that it is possible to select specific
well-behaved forms of the function $f$ such that one of these two
conditions is satisfied for any value of the angular frequency
$\omega$. Also, since the masses only depend on the values of $f$ and
its derivatives at $R=\mathcal R=0$, any higher-order term on $R$ and
$\mathcal R$ up to infinity can be added to the function $f$ leaving
these results unaffected, being then coherent with the two general
constraints we imposed on the function $f$ to begin with.  It would be
of interest to see the restrictions imposed on $f$ by vector and
tensor perturbations of the Kerr solution.

\begin{acknowledgments}
JLR acknowledges Funda\c{c}\~{a}o para a Ci\^{e}ncia e Tecnologia
(FCT)-Portugal and IDPASC for support through grant
no.~PD/BD/114072/2015 and Fulbright Comission Portugal.  JPSL thanks
FCT for partial financial support through Project
no.~PEst-OE/FIS/UI0099/2019.  FSNL acknowledges support from the
Scientific Employment Stimulus contract with reference
CEECIND/04057/2017, and funding from FCT projects
no.~UID/FIS/04434/2019 and no.~PTDC/FIS-OUT/29048/2017.
\end{acknowledgments}

\appendix
 
\section{Scalar-tensor representation of GHMP gravity}
\label{scalarrep}

The objective of this appendix is to show that one can perform the
perturbative analysis in the  scalar-tensor representation
of the GHMP
theory and that the perturbation equations and the results are
the same. The scalar-tensor representation can be achieved by
considering an action with two auxiliary fields, $\alpha$ and $\beta$,
respectively, in the following form
\beq
S=\frac{1}{2\kappa^2}\int_\Omega \sqrt{-g}\Big[f
\left(\alpha,\beta\right)+\frac{\partial f}{\partial \alpha}
\left(R-\alpha\right)
    \nonumber  \\
+\frac{\partial f}{\partial\beta}\left(\cal{R}-\beta\right)
\Big]d^4x+S_m.\label{gensca}
\eeq
Using $\alpha=R$ and $\beta=\mathcal{R}$ we recover the initial action
in Eq.~\eqref{genac}. Therefore, we can define two scalar fields as
$\varphi=\partial f/\partial\alpha$ and $\psi=-\partial
f/\partial\beta$, where the negative sign is set here for
convention. The equivalent action is of the form
\be
S=\frac{1}{2\kappa^2}\int_\Omega \sqrt{-g}
\left[\varphi R-\psi\mathcal{R}-V\left(\varphi,\psi\right)\right]d^4x,
  \label{action3}
\ee
where we defined the potential $V$ as
\be\label{potential}
V\left(\varphi,\psi\right)=-f\left(\alpha,\beta\right)+
\varphi\alpha-\psi\beta.
\ee
We now have an action with four independent variables, namely the
metric $g_{ab}$, the independent connection $\hat\Gamma$, and the
scalar fields $\varphi$ and $\psi$. The equation of motion for
$\hat\Gamma$ remains the same as in the geometrical representation and
it is given by Eq.~\eqref{eqvar1} or, equivalently, by
Eq.~\eqref{riccirel}. Using the definitions of the scalar fields, this
now becomes
\be\label{riccirel2}
\mathcal R_{ab}=R_{ab}-\frac{1}{\psi}
\left(\nabla_a\nabla_b+\frac{1}{2}g_{ab}\Box
\right)\psi+\frac{3}{2\psi^2}\partial_a \psi\partial_b \psi.
\ee
Varying Eq.~\eqref{action3} with respect to
the metric $g_{ab}$ yields the field equation
\beq
&&\varphi R_{ab}-\psi\mathcal R_{ab}-\frac{1}{2}g_{ab}
\left(\varphi R-\psi\mathcal R-V\right)-\nonumber \\
&&-\left(\nabla_a\nabla_b-g_{ab}\Box\right)\varphi=0,
\label{w1}
\eeq
where we used the fact that $T_{ab}=0$ for the solutions
in which we are
interested in this paper. This equation is in agreement with the
geometrical representation in the sense that it can be obtained from
Eq.~\eqref{genfield} simply by using the definitions of the scalar fields
$\varphi$, $\psi$, and the potential $V$. Finally, varying the action
in Eq.~\eqref{action3} with respect to the scalar fields $\varphi$ and
$\psi$ yields directly
\be
R=V_\varphi,\ \ \ \ \ \mathcal R=-V_\psi,
\ee
where the subscripts $\varphi$ and $\psi$ denote derivatives with
respect to the scalar fields $\varphi$ and $\psi$, respectively.

Using Eq.~\eqref{riccirel2} and its trace to cancel the terms
$\mathcal R_{ab}$ and $\mathcal R$ in Eq.~\eqref{action3}, and tracing
the result, one verifies that one of the possible ways for a solution
in general relativity with $R=0$ to be a solution for this
representation of the GHMP
gravity is to impose that $V=0$ and also
that both scalar fields $\varphi$ and $\psi$ are constants. Note that
the trace of the field equation is a PDE for $\varphi$ and $\psi$,
like in Sec.~\ref{sec2} where it was a PDE for $\mathcal R$, and therefore
these solutions are not unique. We choose constant scalar fields as
solutions because this is equivalent to setting $\mathcal R=\mathcal
R_0$ for some constant $\mathcal R_0$, and we recover the results of
the geometrical representation. Then, using $\psi=\psi_0$ for some
constant $\psi_0$ in the trace of Eq.~\eqref{riccirel2} one verifies
that $\mathcal R=R=0$, which is the same result we obtained
before. The constraint $V=0$, for solutions with $\mathcal R=R=0$ is
equivalent to the constraint $f\left(0,0\right)=0$ that we obtained in
Sec.~\ref{sec2}. On the other hand, constraining $\varphi$ and $\psi$
to be constants is equivalent to constraining $f_R$ and $f_\mathcal R$
to be constants in the geometrical representation, which is exactly
what happens for $\mathcal R=R=0$, and thus these results are consistent
with the ones from Sec.~\ref{sec2}.

Now, let us perturb the metric $g_{ab}$ in the same way we did in
Eq.~\eqref{pert}. This will again impose a perturbation in both $R$
and $\mathcal R$ of the forms $R=\bar R+\epsilon\delta R$ and
$\mathcal R=\bar{\mathcal R}+\epsilon\delta\mathcal R$, plus
additional perturbations of the scalar fields of the form
$\varphi=\bar\varphi+\epsilon\delta\varphi$
and $\psi=\bar\psi+\epsilon\delta\psi$.
From the definitions of the scalar fields and using the fact that
$\alpha=R$ and $\beta=\mathcal R$, we can rewrite the perturbations in
the scalar fields as
\be
\delta \varphi=\frac{\partial^2f}{\partial\alpha^2}
\delta\alpha+\frac{\partial^2f}{\partial\alpha\partial
\beta}\delta\beta=\bar f_{RR}\delta R+\bar
f_{R\mathcal R}\delta\mathcal R,
\ee
\be
\delta \psi=\frac{\partial^2f}{\partial\beta^2}\delta\beta+
\frac{\partial^2f}{\partial\beta\partial\alpha}\delta\alpha
=\bar f_{\mathcal R\mathcal R}\delta \mathcal R+\bar f_{\mathcal R R}
\delta R.
\ee
Inserting these perturbations into the traces of
Eqs.~\eqref{riccirel2} and~\eqref{w1} and keeping only
the terms to leading order in
$\epsilon$ yields again the same equations as Eqs.~\eqref{eqsys1} and
\eqref{eqsys2}, and the procedure is the same as that in
Sec.~\ref{sec2}. We therefore conclude that the analysis of metric
perturbations in both representations of the theory is equivalent, as
anticipated.


\end{document}